\documentstyle[aps,prl,twocolumn,floats,epsfig]{revtex}

\begin{document}

\bibliographystyle{prsty}

\title{ 
Uniform susceptibility of classical antiferromagnets in one and two
dimensions \\
in a magnetic field 
\vspace{-1mm}
}      

\author{
  D. Hinzke,\cite{e-hin} U. Nowak\cite{e-now} 
  }

\address{
  Theoretische Tieftemperaturphysik, Gerhard-Mercator-Universit\"at-Duisburg,
  D-47048 Duisburg, Germany}

\author{
  D. A. Garanin\cite{e-gar} 
   }
 
\address{
  Max-Planck-Institut f\"ur Physik komplexer Systeme, N\"othnitzer Strasse 38,
  D-01187 Dresden, Germany\\ 
  \smallskip {\rm(Received 21 April 1999)} \bigskip\\ 
  \parbox{14.2cm}
  {\rm We simulated the field-dependent magnetization $m(H,T)$ and the
    uniform susceptibility $\chi(H,T)$ of  
    classical Heisenberg antiferromagnets in the chain and square-lattice
    geometry using Monte Carlo methods. The results confirm the singular behavior of $\chi(H,T)$ at small $T,H$: $\lim_{T\to
      0}\lim_{H\to 0} \chi(H,T)=1/(2J_0)(1-1/D)$ and $\lim_{H\to
      0}\lim_{T\to 0} \chi(H,T)=1/(2J_0)$, where $D=3$ is the number
    of spin components, $J_0=zJ$, and $z$ is the number of nearest
    neighbors.  A good agreement is achieved in a wide range of
    temperatures $T$ and magnetic fields $H$ with the first-order
    $1/D$ expansion results [D. A. Garanin, J. Stat. Phys. {\bf 83},
    907 (1996)].  \smallskip
    \begin{flushleft}
    \end{flushleft}
    } 
  } 
\maketitle

In recent years, investigations of two-dimensional antiferromagnets
concentrated primarily on the quantum model with $S=1/2$.
A practical reason for that is its possible relevance for the high-temperature
superconductivity.
On the other hand, the identification with the quantum nonlinear sigma model
(QNL$\sigma$M) in the low-energy sector allowed using field-theory
methods \cite{chahalnel8889,hasnie93}.
Although the QNL$\sigma$M results for the $S=1/2$ model proved to be
in a good argeement with quantum Monte Carlo (QMC) simulations (see,
e.g., Ref.\ \cite{kimtro98}), the requirement of low energies confines
the validity region of the QNL$\sigma$M to rather low temperatures
already for $S\geq 1$.
High-temperature series expansions (HTSE) for $S\geq 1$
\cite{elsetal95} and QMC simulations \cite{hartrokaw98} for $S=1$ in
the experimentally relevant temperature range, as well as experiments
on model substances with $1\leq S \leq 5/2$, showed much better accord
with the pure-quantum self-consistent harmonic approximation (PQSCHA)
\cite{cuctogvaiver969798}, than with the
field-theoretical QNL$\sigma$M predictions.
In contrast to the QNL$\sigma$M, the PQSCHA maps a quantum system on
the corresponding classical system on the lattice, which, in turn, can
be studied by classical MC simulations or other methods.
The parameters of these classical Hamiltonians are renormalized by
quantum fluctuations and given by explicit analytical expressions.

The above arguments show that in most cases the classical model can be
used as a good starting point for studying quantum systems.
In fact, most of nontrivial features of two-dimensional
antiferromagnets, such as impossibility of ordering at nonzero
temperatures in the isotropic case, are universal and appear already
at the classical level.
The main theoretical problem is that due to Goldstone modes, a
simple spin-wave theory at $T\ll JS^2$ is inapplicable to
two-dimensional magnets.

Despite their importance, classical antiferromagnets received much
less attention than the quantum $S=1/2$ model.
In particular, the initial uniform susceptibility $\chi(T)$ for the
square lattice having a flat maximum at $T\sim J$ has been simulated
for $S=1/2$ in Refs.\ \cite{okakiknag88,kimtro98} and for $S=1$ in
Ref.\ \cite{hartrokaw98}, but there are no results for the classical
model yet!
For the latter, only the old MC data for the energy \cite{shetob80}
are available up to now.

On the other hand, classical magnets can be theoretically studied with
the help of the $1/D$ expansion, where $D$ is the number of spin
components \cite{abe7273,abehik7377,gar94jsp,gar96jsp}.
In Ref. \cite{gar94jsp}, $\chi(T)$ has been calculated for the square lattice
and linear chain to first
order in $1/D$ for all temperatures, the solution interpolating
between the exact result at $T=0$ and the leading terms of the HTSE at
high temperatures.
In contrast, the low-energy approaches such as ``Schwinger-boson
mean-field theory'' \cite{auearo88prlprb} or ``modified
spin-wave theory'' \cite{tak8789} break down at $T\gtrsim J$ and fail
to reproduce the maximum of $\chi(T)$.
It should be noted that for quantum magnets there is a method
consisting in the expansion in powers of $1/N$ where $N$ is the number
of flavors in the Schwinger-boson technique \cite{timgirhensan98}.
This method, which is {\em nonequivalent} to the $1/D$ expansion in
the limit $S\to\infty$, is supposed to work for all $T$, in contrast
to the low-energy QNL$\sigma$M.
Unfortunately, only the results for $m(T,H)$ of {\em ferromagnets}
\cite{timgirhensan98} are available.

The $1/D$ expansion also works in the situations with nonzero
magnetic field, which are not amenable to the methods of Refs.\ 
\cite{auearo88prlprb,tak8789} imposing an external
condition $m=0$.
An especially interesting issue is the singular behavior of
$\chi(H,T)$ for $H,T\to 0$ for the square-lattice and linear-chain models.
For any $H\ne 0$, the spins with lowering temperature come into a
position nearly perpendicular to the field, thus $\lim_{H\to 0}\lim_{T\to 0}
\chi(H,T)=1/(2J_0)$, where $J_0$ is the zero Fourier transform of the
exchange interaction, $J_0=zJ$, $z$ is the number of nearest
neighbors.
This value coincides with the susceptibility of the three-dimensional classical
antiferromagnets on bipartite lattices in the direction transverse to the spontaneous magnetization. 
For $H=0$, the spins assume {\em all} directions, including that along the
infinitesimal field, for which the susceptibility tends to zero at $T\to 0$.
Thus $\lim_{T\to 0}\lim_{H\to 0} \chi(H,T)=1/(2J_0)(1-1/D)$.
One can see that the difference between these two results is captured
{\em exactly} in the first order in $1/D$.
According to Ref.\ \cite{gar96jsp}, for any $H\ne 0$
with lowering temperature $\chi(H,T)$ increases, goes through the flat
maximum, decreases, attains a minimum and then goes up to the limiting
value $1/(2J_0)$.

The existence of the interesting features described above, which should be
also pertinent to quantum antiferromagnets, have never been checked
numerically. That is why we have undertaken MC simulations for
classical AFMs in square-lattice and linear-chain geometries.

Our systems are defined by a classical Heisenberg Hamiltonian
\begin{equation}
  {\cal H} = -{\bf H}\sum_{i}{\bf S}_i +
  \frac{1}{2}\sum_{ij}J_{ij}{\bf S}_i{\bf S}_j
\label{e:DHam}
\end{equation}
where ${\bf S}$ is a $D$-component normalized vector of unit length
($|{\bf S}|=1$), ${\bf H}$ is a magnetic field and the exchange coupling
$J_{ij}$ is $J>0$ for nearest neighbors and zero otherwise. 
The mean-field transition temperature is given by $T_c^{\rm MFA}=J_0/D=zJ/D$.
Although there is no phase transition in our model, it is convenient to choose 
$T_c^{\rm MFA}$ as the energy
scale and to introduce dimensionless temperature, magnetic field, and
susceptibilities

\begin{equation}
  \theta\equiv T/T_c^{\rm MFA}, \qquad h\equiv H/J_0, \qquad \tilde
  \chi_\alpha \equiv J_0 \chi_\alpha ,
\label{e:RedVars}
\end{equation}
where $\chi_\alpha \equiv \partial\langle S_\alpha\rangle/\partial H_\alpha$
and $\alpha=x,y,z$.

In the limit $D\to\infty$, the model Eq. (\ref{e:DHam}) is exactly
solvable and equivalent to the spherical model.
The solution includes an integral over the Brillouin zone taking into
account spin-wave effects in a nonperturbative way.
The latter leads to the absence of the phase transition for the
spatial dimensionalities $d \leq 2$.

The $1/D$ corrections to the spherical-model solution have been obtained in
Refs.\ \cite{abe7273,abehik7377,gar94jsp,gar96jsp}.
They include double integrals over the Brillouin zone and are
responsible for the maximum of the antiferromagnetic susceptibility at
$\theta\sim 1$ \cite{gar94jsp}.  For small fields and temperatures,
$h,\theta \ll 1$, the field-induced magnetization $m$ for the square-lattice model
simplifies to
\begin{equation}
  m\cong \frac h2 \left[ 1 - \frac 1D + \frac{ \theta }{ \pi D } \ln \left( 1 +
      \frac{ h^2 }{ 16 } e^{\pi/\theta} \right) + \frac \theta D \right],
  \label{e:mLowht}
\end{equation}
which follows from Eqs.\ (4.9) and (2.23) of Ref.\ \cite{gar96jsp}.
The log term of the above expression is responsible for the
singularity of both transverse and longitudinal (with respect to the field) 
susceptibilities,

\begin{equation}
  \tilde\chi_\perp \equiv m/h, \qquad \tilde\chi_\| \equiv \partial
  m/\partial h,
\label{e:chiDef}
\end{equation}
which was mentioned above.
For $h=0$ they have the form $\tilde\chi \cong [1 - 1/D +
\theta/D]/2$, whereas for $h\neq 0$ the limiting value at $\theta=0$
and the slope with respect to $\theta$ are different: $\tilde\chi
\cong \{1 - [\theta/(\pi D)]\ln[16/(e^\pi h^2)]\}/2$.
In the latter case, $\chi$ has a minimum at $\theta \cong \theta^* =
\pi/\ln(16/h^2)$.
There are corrections of order $\theta^2$ and $1/D^2$ to Eq.\ 
(\ref{e:mLowht}).
The latter renormalize the last, regular term in Eq.\ (\ref{e:mLowht})
(see Eq.\ (8.2) of Ref.\ \cite{gar94jsp}).
The $1/D^2$ corrections cannot, however, appear in the log term of
Eq.\ (\ref{e:mLowht}), because this would violate the general properties
of $\chi(H,T)$ discussed above.

For the linear chain, the magnetization in the region $h,\theta \ll 1$ to
first order in $1/D$ is given by \cite{gar96jsp}
\begin{equation}\label{e:mLowht1d}
m \cong \frac{h}{2}
\left[
1 -\frac{\theta}{D\sqrt{h^2+\theta^2}} + \frac{ \theta }{ D } + O(\theta^2)
\right].
\end{equation}
The transverse susceptibility of the linear chain behaves qualitatively
similarly to that of the square lattice.
The minimum of $\chi_\perp$ is attained at $\theta=h^{2/3}$ which is smaller
than in two dimensions.
The longitudinal susceptibility $\chi_\|$ corresponding to Eq.\ (\ref{e:mLowht1d}) has a 
minimum at $\theta \cong 3^{1/3}h^{2/3}\gg h$ and a {\em maximum} at
$\theta \cong 3^{-1/2}h^{3/2} \ll h$.

For comparison, the zero-field Takahashi's results \cite{tak8789} for the
Heisenberg model on the linear chain and square lattice can for $\theta \ll 1$
be rewritten in the form \cite{gar94jsp}
\begin{equation}\label{Tak}
\tilde\chi \cong \frac 13
\left\{
\begin{array}{ll}
\displaystyle
[1-\theta/3]^{-1},           &  d=1\\
\displaystyle
2\left[1+\sqrt{1-(4/3)\theta}\right]^{-1},        & d=2, 
\end{array}
\right.
\end{equation}
where the exponentially small terms are neglected.
For both lattices the low-temperature expansion is the same to order $\theta$:
$\tilde\chi = (1/3) + (1/9)\theta + \ldots$, and the results diverge at
$\theta\sim 1$.
The coefficient in front of $\theta$ here is at variance with the
$1/D$-expansion results above for $D=3$.
It was argued in Ref. \cite{gar94jsp} that the correct general-$D$ form of the
low-temperature expansion of the zero-field susceptibility for both square
lattice and the linear chain reads
\begin{equation}\label{chiLowT}
\tilde\chi = \frac 12
\left(1-\frac 1D \right) + \frac{1}{2D} \left(1-\frac 1D \right)\theta 
 + O(\theta^2),
\end{equation}
i.e., it is reproduced to order $\theta$ at the {\em second} order of the
$1/D$ expansion.
This formula is in accord with Takahashi's theory.


In order to check the validity of the analytic results from the $1/D$
expansion above for the most realistic case of $D=3$, we performed Monte
Carlo simulation for three-component classical spins on a chain with
length $N$ as well as on a square lattice of size $N = L \times L$,
both with periodic boundary conditions. 
In our Monte Carlo procedure, a spin is chosen randomly and a trial
step is made where the new spin direction is taken randomly with equal
distribution on the unit sphere. 
This trial step does not depend on
the initial spin direction.  
The energy change of the system is
computed according Eq.\ (\ref{e:DHam}) and is accepted with the heat-bath
probability. 
One sweep through the lattice and performing the
procedure described above once per spin (on average) is called one
Monte Carlo step (MCS). We start our simulation at high temperature
and cool the system stepwise. 
For each temperature we wait 6000MCS
(chain) and 4000MCS (square lattice), respectively, in order to reach
equilibrium.  
After thermalization we compute thermal averages
$\langle \ldots \rangle$ for the next 8000MCS (chain) and 6000MCS
(square lattice), respectively.

The relevant quantities we are interested in are 
 the magnetization $m \equiv m_z = \langle M_z \rangle$
and the components of the susceptibility
 $\chi_\alpha = \frac{N}{T} (\langle M_\alpha^2 \rangle -
\langle M_\alpha \rangle^2)$, where the $z$ axis is directed along {\bf H},  
$\alpha = x,y,z$, and $M_\alpha \equiv \frac{1}{N} \sum_i S_i^\alpha$.
We have used the formula above for $\chi_\alpha$ to simulate the zero-field and
longitudinal susceptibility, $\chi_\| \equiv \chi_z$.
For the transverse susceptibility, $\chi_\perp \equiv \chi_x=\chi_y$, at nonzero
field it is more convenient to use Eq.\ (\ref{e:chiDef}).
For $h = 0$ the transverse and  longitudinal susceptibilities are identical
and calculated as $\chi_\perp = \chi_\| = (\chi_x+\chi_y+\chi_z)/3$.

With intent to minimize the statistical error and to be able to
compute error bars we take averages over $N_r=100$ independent Monte Carlo
runs. The error bars we show are the mean errors of the averages
$\sigma/\sqrt{N_r}$, where $\sigma$ is the standard deviation
of the distribution of thermal averages following from the independent
runs.
\begin{figure}[t]      
  \unitlength1cm
  \begin{picture}(8,6)(0,-7)
    \centerline{\epsfig{file=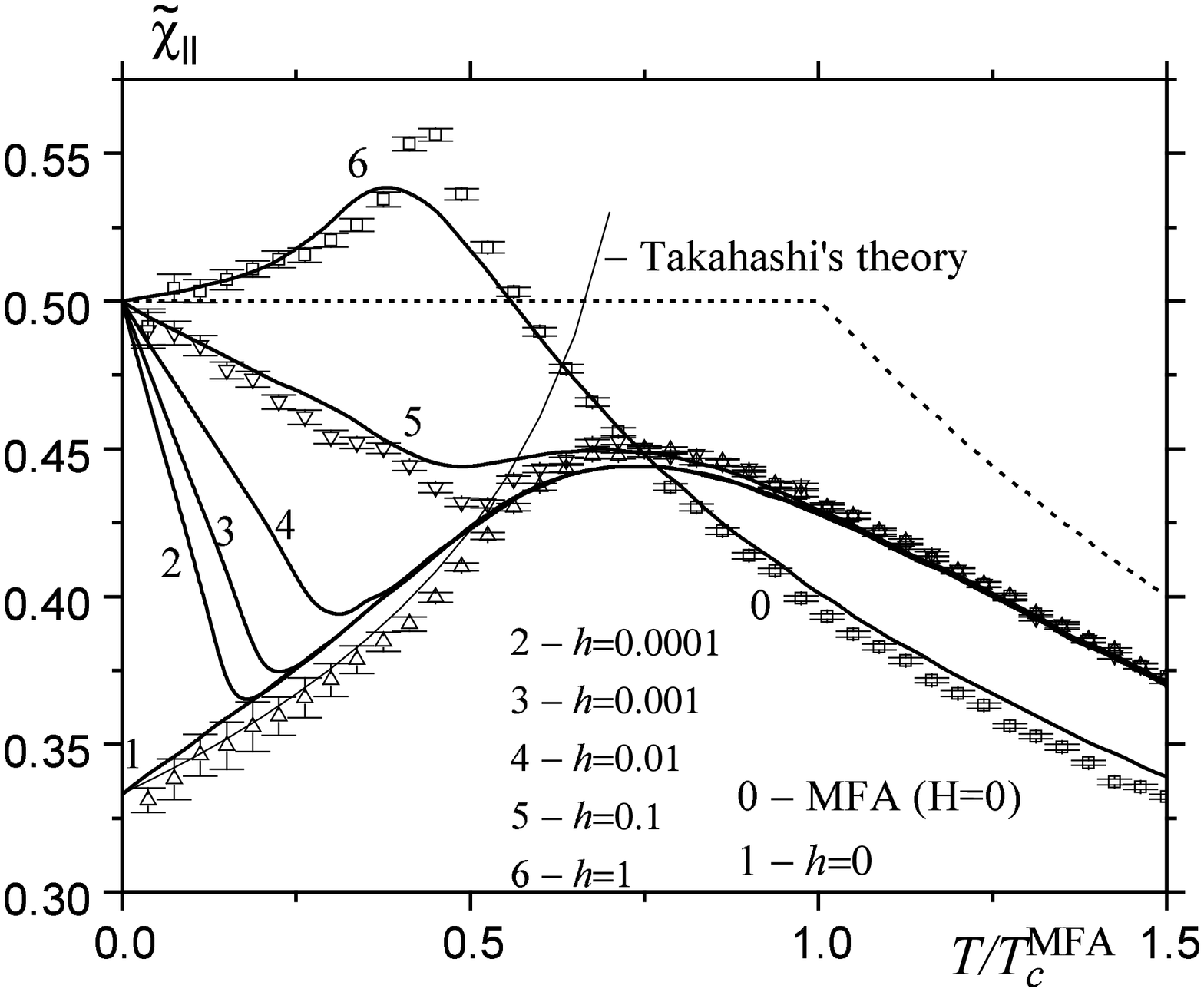,angle=-90,width=8cm}}
  \end{picture}
%
  \begin{picture}(8,6)(0,-7)
    \centerline{\epsfig{file=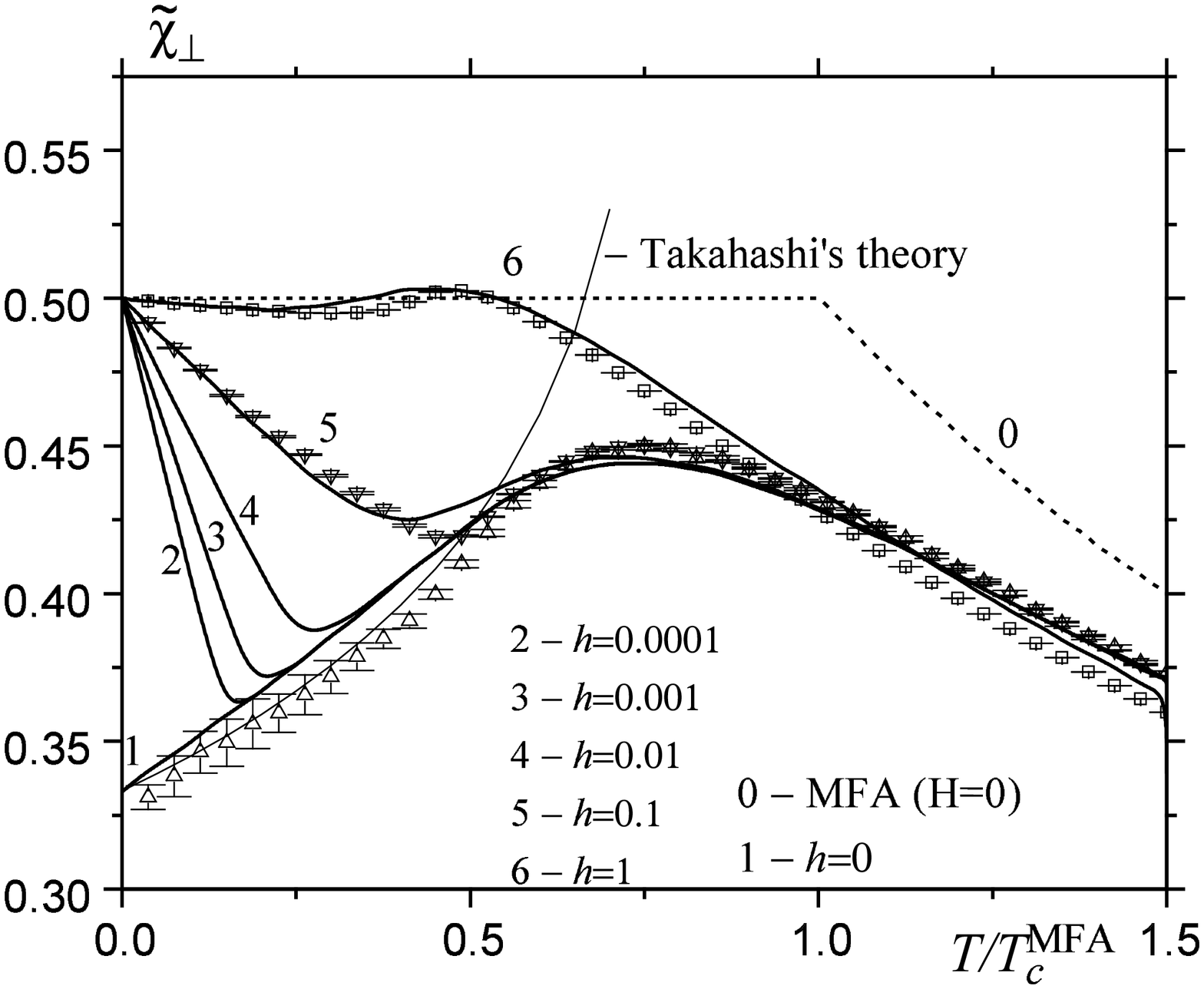,angle=-90,width=8cm}}
  \end{picture}
    \caption{Temperature dependence of the longitudinal and transverse susceptibility for
      the square lattice for different values of the magnetic field
      $h$. The points are results from Monte Carlo simulations for $L
      = 64$ and $h=0, 0.1$, and 1.
  \label{f:chi2d}
  }  
\end{figure}

\begin{figure}[t]      
  \unitlength1cm
  \begin{picture}(8,6)(0,-7)
    \centerline{\epsfig{file=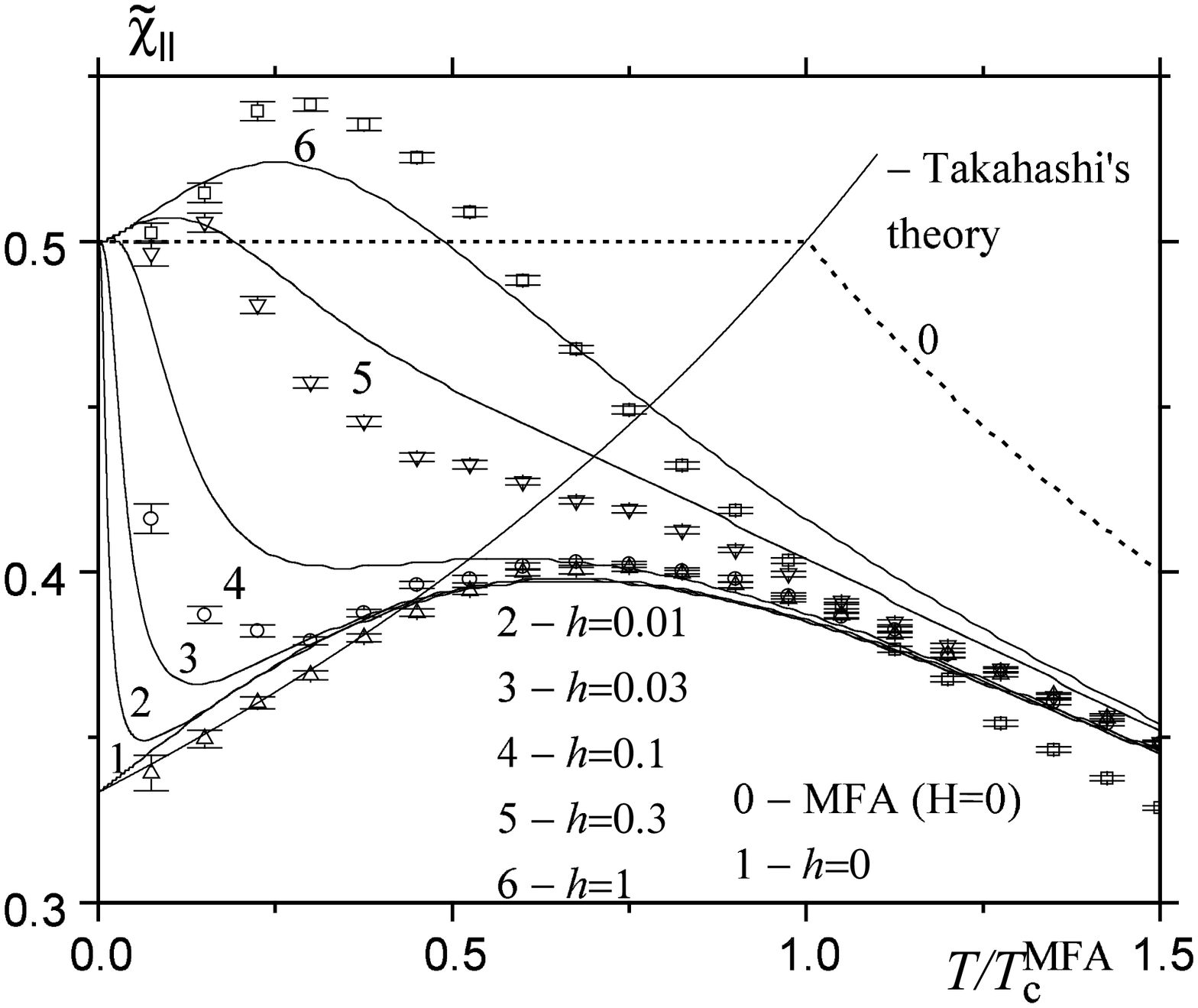,angle=-90,width=8cm}}
  \end{picture}
%
  \begin{picture}(8,6)(0,-7)
    \centerline{\epsfig{file=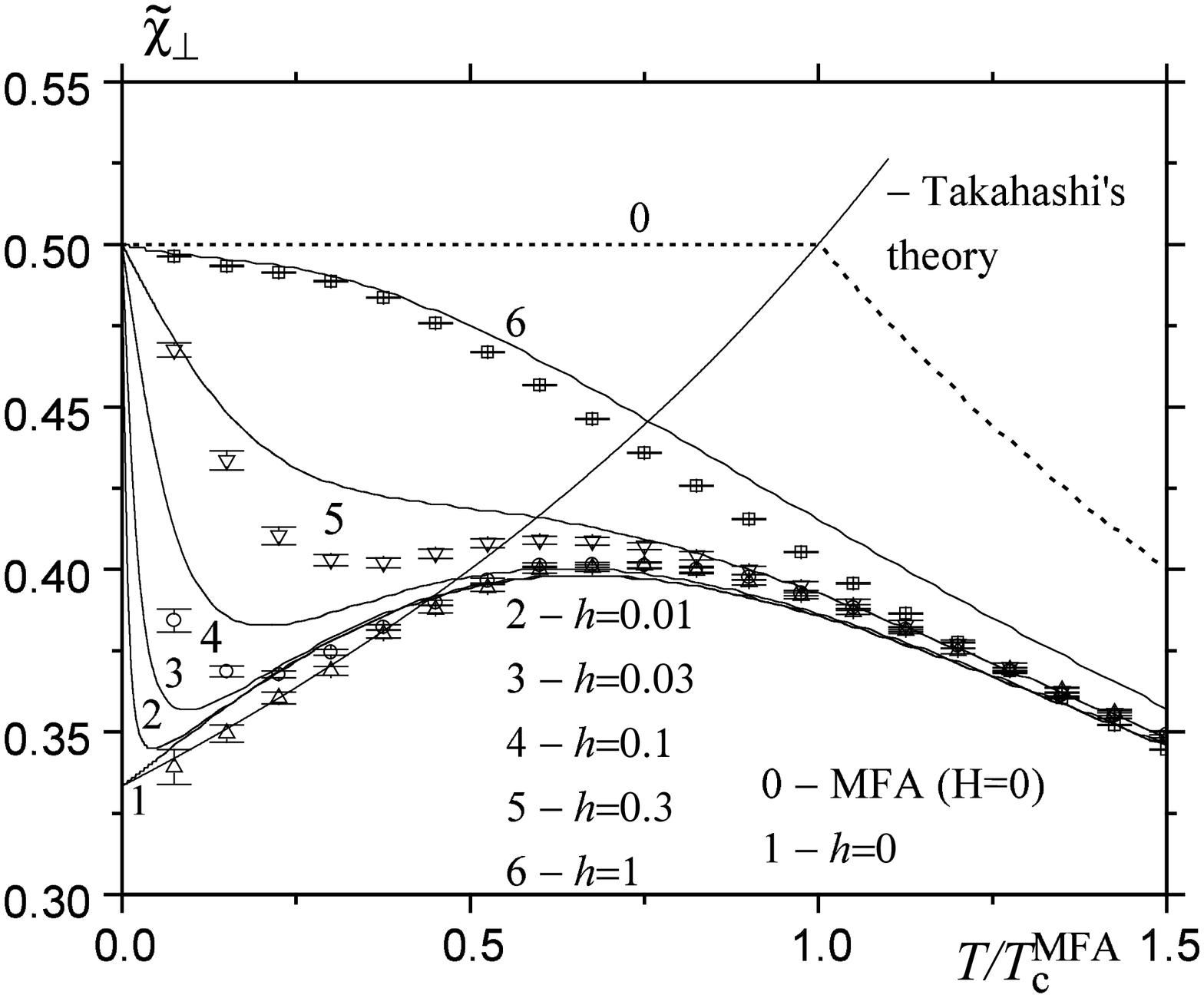,angle=-90,width=8cm}}
  \end{picture}
    \caption{Temperature dependence of the longitudinal and transverse susceptibility for
      the chain for different values of the magnetic field. The points
      are results from Monte Carlo simulations for $L = 100$ and
      $h=0, 0.1, 0.3$, and 1.
    \label{f:chi1d}
  }  
\end{figure}

We start the comparison of theoretical and numerical results with the
square lattice.  Fig.~\ref{f:chi2d} shows the temperature dependence
of the reduced longitudinal susceptibility $\tilde\chi_\|$ and 
reduced transverse susceptibility $\tilde\chi_\perp$ for different values of the
magnetic field, both for the system size $L = 64$. The corresponding results for the spin chain with system size $L = 100$ are presented in Fig.~\ref{f:chi1d}.

We investigated possible finite-size effects by varying the lattice
size. However, we did not find any significant change of our data for
lattice sizes in the range $L = 16 \ldots 64$ (square lattice) and $L
= 40 \ldots 100$ (linear chain). Also, we did not find any systematic
change of our results for longer Monte Carlo runs so that we believe
to present data corresponding to thermal equilibrium.

Note, that for all Monte Carlo data shown the error bars of the
transverse susceptibility are smaller than those of the longitudinal
one since the transverse susceptibility follows directly from the $z$
component of the magnetization while the longitudinal susceptibility
is calculated from the {\em fluctuations} of the $z$ component of the
magnetization.  In the case $h = 0$ the transverse and longitudinal
susceptibility are identical and follow from fluctuations of the
magnetization so that the error bars are larger.


For the square lattice as well as for the chain the numerical data
confirm the non-analytic behavior of $\chi$ in the limit of
temperature $T \to 0$, i.~e. the limiting values $\tilde\chi_\perp = \tilde\chi_\|
= 1/2$ for $h \ne 0$ and $\tilde\chi_\perp = \tilde\chi_\| = 1/3$ for $h = 0$.

Especially for the square lattice, the Monte Carlo data agree
reasonable with the first-order $1/D$ expansion in the whole range of temperatures. 
On the other hand, at low temperatures the agreement with Takahashi's theory
within error bars is achieved.
Our numerical data thus confirm that the coefficient in the linear-$\theta$
term in $\chi$ in Takahashi's theory is accurate. 
For $h=1$ and $\theta \gtrsim 1$, the MC data fall slightly below the 
$1/D$-expansion curve.
Both are again in accord with each other for $\theta \gtrsim 3$ (not shown).

The maximum of the longitudinal susceptibility of the square-lattice model 
for $h=1$ looks much sharper than that of the theoretical curve. 
This feature,
as well as the hump on the $h=0.1$ curve at slightly lower temperature, are
possible indications of the Berezinsky-Kosterlitz-Thouless (BKT) transition.
The reason for that is an effective reduction of the number of spin 
components by one at sufficiently low temperatures in the magnetic field
(the effect mentioned in the introduction), so that the Heisenberg model becomes effectively $D=2$
and it can undergo a BKT transition in two dimensions.
We have not, however, studied this point in detail in this work.

For the antiferromagnetic chain our MC simulation data are in a qualitative 
agreement with the $1/D$ expansion, although the discrepancies are stronger.

Unfortunately, we could also not perform simulations for even lower
values of the field $h$ for the following reason: The singular
behavior of $\chi$ stems from the fact that for $h > 0$ the spins tend
to come into a position perpendicular to the field.  For fields as small as
$ h = 0.01$ (curve 4 in Figures 1 and 2) the amount of energy related
to this ordering field is 100 times smaller than the exchange
interaction energy. Therefore the corresponding relaxation for this
energetically favorable state takes very long in a Monte Carlo
simulation, especially for these low temperatures, where this effect
occurs for low fields.

Our MC simulations showed for the first time the singular behavior of the
susceptibility of classical antiferromagnets at low temperatured and magnetic fields.
The results are in accord with predictions based on the first-order $1/D$
expansion \cite{gar94jsp,gar96jsp}. 
It would be interesting to try deriving the corresponding low-temperature
results [cf. Eqs.\ (\ref{e:mLowht}) and (\ref{e:mLowht1d})] without using the
$1/D$ expansion.
One of the formulas of this type already exists: It is Eq.\ (\ref{chiLowT}).
A candidate among theoretical approaches is the chiral perturbation theory of
Ref.\ \cite{hasnie93}, which is applicable to quantum models, as well.

The features manifested here by classical antiferromagnets should be pertinent
to quantum models, as well.
The effects observed here could be checked with the help of the QMC
simulations which achieved recently a substantial accuracy (see,
e.g., Refs.\ \cite{kimtro98} and \cite{hartrokaw98}).
Another possibility is to map the quantum model on the
classical one \cite{cuctogvaiver969798} and to
perform classical MC simulations. 
One should also mention an alternative way of mapping of quantum magnetic
Hamiltonians on
classical ones with the help of the coherent-state cumulant expansion 
\cite{klafulgar99epl}, which is a rigorous expansion in powers of $1/S$.

\vspace{-0.5cm}


\end{document}